\documentclass[aps,prl,superscriptaddress,twocolumn]{revtex4-1}
\usepackage{graphicx}
\usepackage[dvipsnames]{xcolor}
\usepackage{siunitx}
\usepackage{hyperref}
\usepackage{xfrac}
\usepackage{soul}
\usepackage{color}
\usepackage{stmaryrd}
\usepackage[nameinlink,capitalise]{cleveref}
\hypersetup{
colorlinks=true,
linkcolor=blue,
filecolor=magenta,
urlcolor=cyan,
}

\makeatletter
\AtBeginDocument{\let\LS@rot\@undefined}
\makeatother
\begin{document}
\title{Measurement of the \texorpdfstring{$^{229}$}{229}Th isomer energy with a magnetic micro-calorimeter}
\author{Tomas Sikorsky}
\thanks{T.S. and J.G. contributed equally to this work.}
\affiliation{Kirchhoff-Institute for Physics, Heidelberg University, INF 227, 69120 Heidelberg, Germany}
\affiliation{Institute for Atomic and Subatomic Physics, TU Wien, Stadionallee 2, 1020, Vienna, Austria}
\author{Jeschua Geist}
\thanks{T.S. and J.G. contributed equally to this work.}
\author{Daniel Hengstler}
\author{Sebastian Kempf}
\author{Loredana Gastaldo}
\author{Christian Enss}
\affiliation{Kirchhoff-Institute for Physics, Heidelberg University, INF 227, 69120 Heidelberg, Germany}
\author{Christoph Mokry}
\affiliation{Johannes Gutenberg University, 55099 Mainz, Germany.}
\affiliation{Helmholtz Institute Mainz, 55099 Mainz, Germany.}
\author{J\"org Runke}
\affiliation{Johannes Gutenberg University, 55099 Mainz, Germany.}
\affiliation{GSI Helmholtzzentrum f\"ur Schwerionenforschung GmbH, 64291 Darmstadt, Germany}
\author{Christoph E. D\"ullmann}
\affiliation{Johannes Gutenberg University, 55099 Mainz, Germany.}
\affiliation{Helmholtz Institute Mainz, 55099 Mainz, Germany.}
\affiliation{GSI Helmholtzzentrum f\"ur Schwerionenforschung GmbH, 64291 Darmstadt, Germany}
\author{Peter Wobrauschek}
\author{Kjeld Beeks}
\author{Veronika Rosecker}
\author{Johannes H. Sterba}
\author{Georgy Kazakov}
\author{Thorsten Schumm}
\affiliation{Institute for Atomic and Subatomic Physics, TU Wien, Stadionallee 2, 1020, Vienna, Austria}
\author{Andreas Fleischmann}
\affiliation{Kirchhoff-Institute for Physics, Heidelberg University, INF 227, 69120 Heidelberg, Germany}
\date{\today}
\begin{abstract}
We present a measurement of the low-energy (0--60\,keV) $\gamma$ ray spectrum produced in the $\alpha$-decay of $^{233}$U using a dedicated cryogenic magnetic micro-calorimeter. The energy resolution of $\sim$\SI{10}{\electronvolt}, together with exceptional gain linearity, allow us to measure the energy of the low-lying isomeric state in $^{229}$Th using four complementary evaluation schemes. The most accurate scheme determines the $^{229}$Th isomer energy to be \SI{8.10\pm0.17}{\electronvolt}, corresponding to \SI{153.1\pm3.7}{\nano\meter}, superseding in precision previous values based on $\gamma$ spectroscopy, and agreeing with a recent measurement based on internal conversion electrons. We also measure branching ratios of the relevant excited states to be $b_{29}=9.3(6)\%$ and $b_{42}=0.3(3)\%$.

\end{abstract}
\maketitle
The low-energy metastable isomeric state in $^{229}$Th ($^{229m}$Th) has fascinated researchers over the past 40 years~\cite{Thirolf_2019}. It is expected to have an excitation energy of $\sim$\SI{8}{\electronvolt}, making it the only nuclear state accessible to laser manipulation known so far. Optical excitation of the $^{229}$Th nucleus would allow to transfer the precision of laser spectroscopy to nuclear structure analysis~\cite{Peik2015}. A vast plethora of applications and investigations have been proposed for the $^{229m}$Th state, ranging from a nuclear gamma laser~\cite{Tkalya11}, highly accurate and stable ion nuclear clock~\cite{Peik2003,Campbell2012} to compact solid-state nuclear clocks~\cite{Kazakov2012}. Such clocks would allow to attain a new level of precision for probes of fundamental physics, e.g., a variation of fundamental constants~\cite{Flambaum06,Thirolf19}, search for dark matter~\cite{Wcislo2016,Arvanitaki2015} or as a gravitational wave detector~\cite{Derevianko2014}. They can be used in different applications, such as geodesy~\cite{Bondarescu15} or satellite-based navigation~\cite{GNSS}.

Despite considerable efforts, neither the resonant optical excitation of $^{229m}$Th from the nuclear ground state nor the emission of fluorescence photons in radiative decay has been observed~\cite{von2018towards}. Several recent attempts to excite the nucleus using broadband synchrotron radiation failed to detect a signal~\cite{Jeet2015,Yamaguchi2015,Stellmer2018}. All currently available information about the existence~\cite{VonderWense2016}, the energy~\cite{Seiferle2019,Beck2007,Beck09,Yamaguchi2019}, or lifetime~\cite{Seiferle17} of the isomer is derived from experiments where the $^{229}$Th isomer is produced in $\alpha$-decay of $^{233}$U, or through the \mbox{x ray} pumping of the second excited nuclear state~\cite{Masuda2019}.

The existence of a low-lying isomeric state in $^{229}$Th was deduced in 1976 from analysis of a $\gamma$ ray spectrum associated with $\alpha$-decay of $^{233}$U~\cite{Kroger76}. Refined measurements of the same spectrum performed in the early nineties with different Ge and Si(Li) detectors, whose energy resolution was about several hundreds of eV, determined the isomer energy to be \SI{3.5\pm1}{\electronvolt}~\cite{Helmer1994}. Reanalysis of these spectra gave \SI{5.5\pm1}{\electronvolt}~\cite{Filho05}. Later, a more precise measurement using a NASA \mbox{x ray} micro-calorimeter spectrometer with an energy resolution of $\sim$\SI{30}{\electronvolt} measured an isomer energy of \SI{7.8\pm0.5}{\electronvolt}, shifting the transition into the vacuum ultraviolet region~\cite{Beck2007,Beck09}. Recently, another calorimetric experiment with $\sim$\SI{40}{\electronvolt} energy resolution reported the isomer energy to be \SI{8.30\pm0.92}{\electronvolt}~\cite{Yamaguchi2019}, consistent with the previous result but not improving upon the uncertainty.

The $^{229m}$Th state was also studied by direct spectroscopy of recoil ions emerging from a $^{233}$U source. The $^{233}$U$\rightarrow^{229}$Th decay populates the $^{229m}$Th state with a 2\% probability~\cite{Thielking2018}. Thorium ions were slowed down in a buffer gas and selectively extracted in a quadrupole mass separator~\cite{VonderWense2016}. Nuclear magnetic dipole and electric quadrupole momenta of the $^{229m}$Th isomer have been deduced from laser spectroscopy data~\cite{Thielking2018}. The lifetime of the $^{229m}$Th state for atoms deposited onto the MCP detector surface has been measured~\cite{Seiferle17}. Finally, an isomer energy of \SI{8.28\pm0.17}{\electronvolt} has been determined by spectroscopy of the internal conversion electrons emitted during the decay of the $^{229m}$Th ions, neutralized by a graphene foil~\cite{Seiferle2019}. This value was deduced by combining spectroscopy data of conversion electrons with calculated distributions of initial/final electronic states. The calculated uncertainty of the initial/final electronic state distribution significantly contributes to the uncertainty of the reported isomer energy (\SI{0.16}{\electronvolt}).

In the work presented here, we perform $\gamma$ spectroscopy using the magnetic micro-calorimeter \mbox{maXs-30}. It is specially designed for optimal performance around \SI{30}{\kilo\electronvolt}, corresponding to the $\gamma$ rays produced in the $^{233}$U$\rightarrow^{229}$Th $\alpha$-decay. This experiment complements the conversion electron experiment in that the isomer energy is extracted directly from the experimental data, without resorting to calculations. The only significant uncertainty in our experiment is the statistical error.

\begin{figure}
\includegraphics[width=\linewidth]{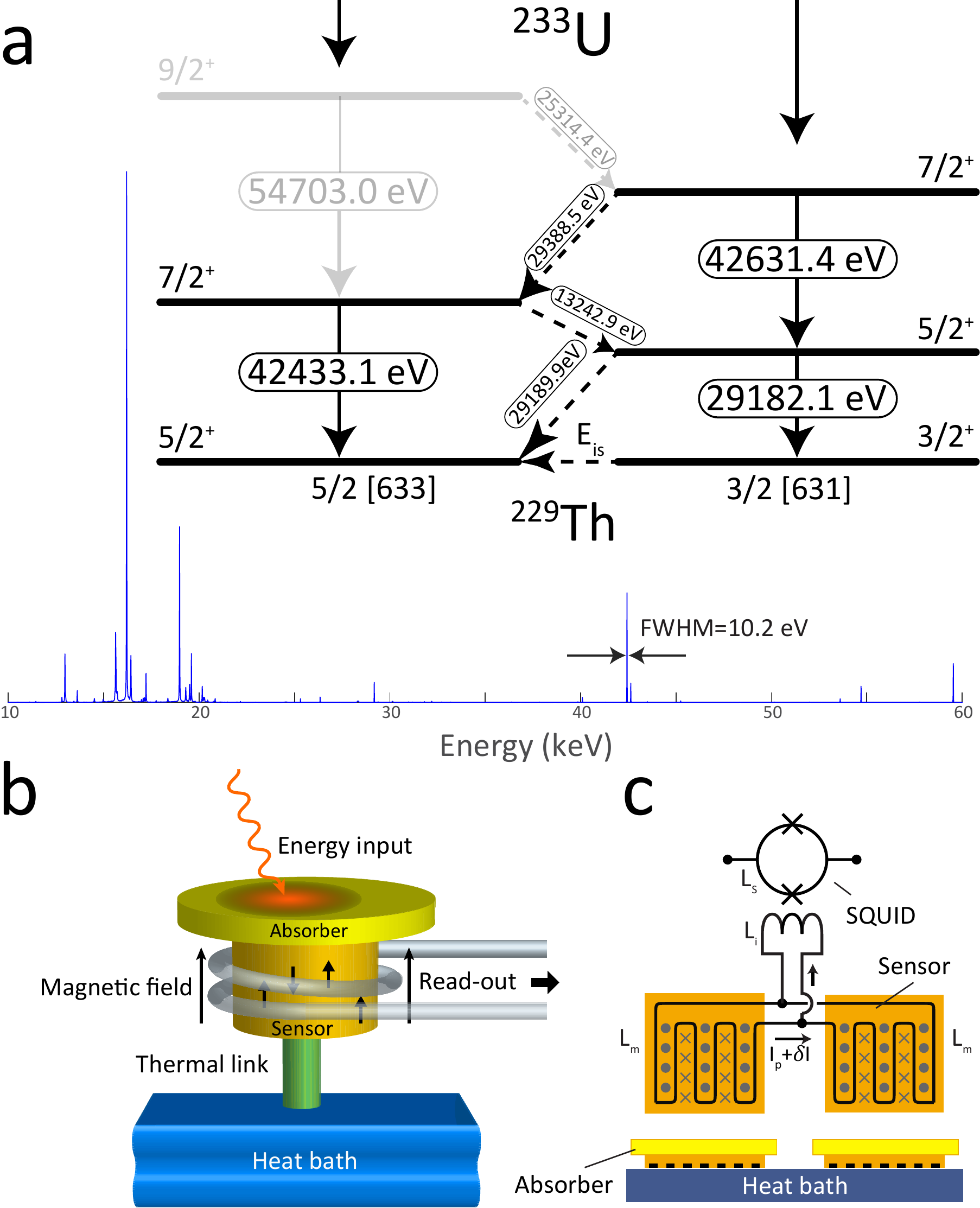}
\caption{(a) Partial nuclear level scheme of the $^{229}$Th nucleus with relevant decay paths and energies. The total spectrum in the energy range up to \SI{60}{\kilo\electronvolt} is shown in the bottom. (b) Schematic of a magnetic micro-calorimeter. A $\gamma$ ray is absorbed in a gold absorber. The heat is then transferred and measured by a Ag:Er$^{3+}$ paramagnetic sensor. A weak thermal link to the heat bath enables thermalization. (c) A persistent current $(I_p)$ circulates in the superconducting meander-shaped pick-up coil polarizing the magnetic moment in the sensor. As the flux in a superconducting loop is conserved, a change of flux $\Delta\Phi$ driven by a temperature-induced change of magnetization induces an additional screening current ($\delta I$), which is readout as a voltage drop over the dc-SQUID.}
\label{FIG. 1.}
\end{figure}

In magnetic micro-calorimeters (MMC), the energy of a $\gamma$ ray is converted into heat in a thin absorber plate (see \cref{FIG. 1.}b). The absorbers used in this experiment are made of \SI{20}{\micro\meter} thick gold layers, realizing 65\,\% stopping efficiency at \SI{30}{\kilo\electronvolt} while having \SI{10}{\electronvolt} resolution. For a precise determination of the small temperature rise on the order of some hundred \SI{}{\micro\kelvin}, MMC's make use of a paramagnetic temperature sensor operated in a weak magnetic field~\cite{Fleischmann2005,Fleischmann2009}. As a paramagnetic sensor, we use silver doped with a few hundred ppm of erbium.

The detector is composed of 8\,$\times$\,8 pixels operated in pairs. Each pair of pixels consists of two gold absorbers, each readout by a Ag:Er$^{3+}$ temperature sensor. Two parallel meander-shaped pick-up coils made of niobium are connected to the input coil of a dc-SQUID current sensor. The pick-up coils of the two sensors generate opposing currents for an equivalent magnetization change in the two sensors. The resulting screening current ($\delta I$) in the input coil of the dc-SQUID is directly proportional to the temperature difference of the two sensors. We read out the screening current as a linearized voltage drop over the dc-SQUID (see \cref{FIG. 1.}c)~\cite{Kempf2017}.

The $\gamma$ radiation is emitted from a solution containing $^{233}$U. Uranium is dissolved in an aqueous solution as uranyl nitrate UO$_2$(NO$_3$)$_2$ and is contained inside a PEEK-capsule. The wall thickness of 2\,mm shields $\alpha$ and $\beta$ radiation but is transparent for photons above few keV. The activity of the source was 74\,MBq, it was chemically purified at the Institute for Nuclear Chemistry, Johannes Gutenberg University Mainz, to remove daughter products of the uranium chain (Th, Ra, Ac) which increase the activity and hence background in the measurement. $\alpha$ and $\gamma$ spectroscopy (using Ge detectors) performed on the source indicate a 2\,$\%$ $^{234}$U and a $<$\,1\,ppm $^{232}$U contamination. Additionally, traces of $^{238,240,242}$Pu, $^{241}$Am, $^{237}$Np where identified (see supplementary material).

The $\gamma$ spectrum in the energy range 0--\SI{60}{\kilo\electronvolt} was recorded over about 640 pixel$\times$days, which corresponds to about 8 million events. For each absorption event, we recorded the full pulse shape, which shows a \SI{9.6}{\micro\second} fast voltage increase, followed by a $\tau=\SI{2.7}{\milli\second}$ decay (see supplementary material). After every event, an electronic hold-off of 450\,ms is used to allow for the thermalization of the detector and avoid pile-ups. A single pulse amplitude value $U$ is extracted for each event by fitting a generic, amplitude-scaled pulse shape~\cite{Enss2000}.

\begin{figure}
\includegraphics[width=\linewidth]{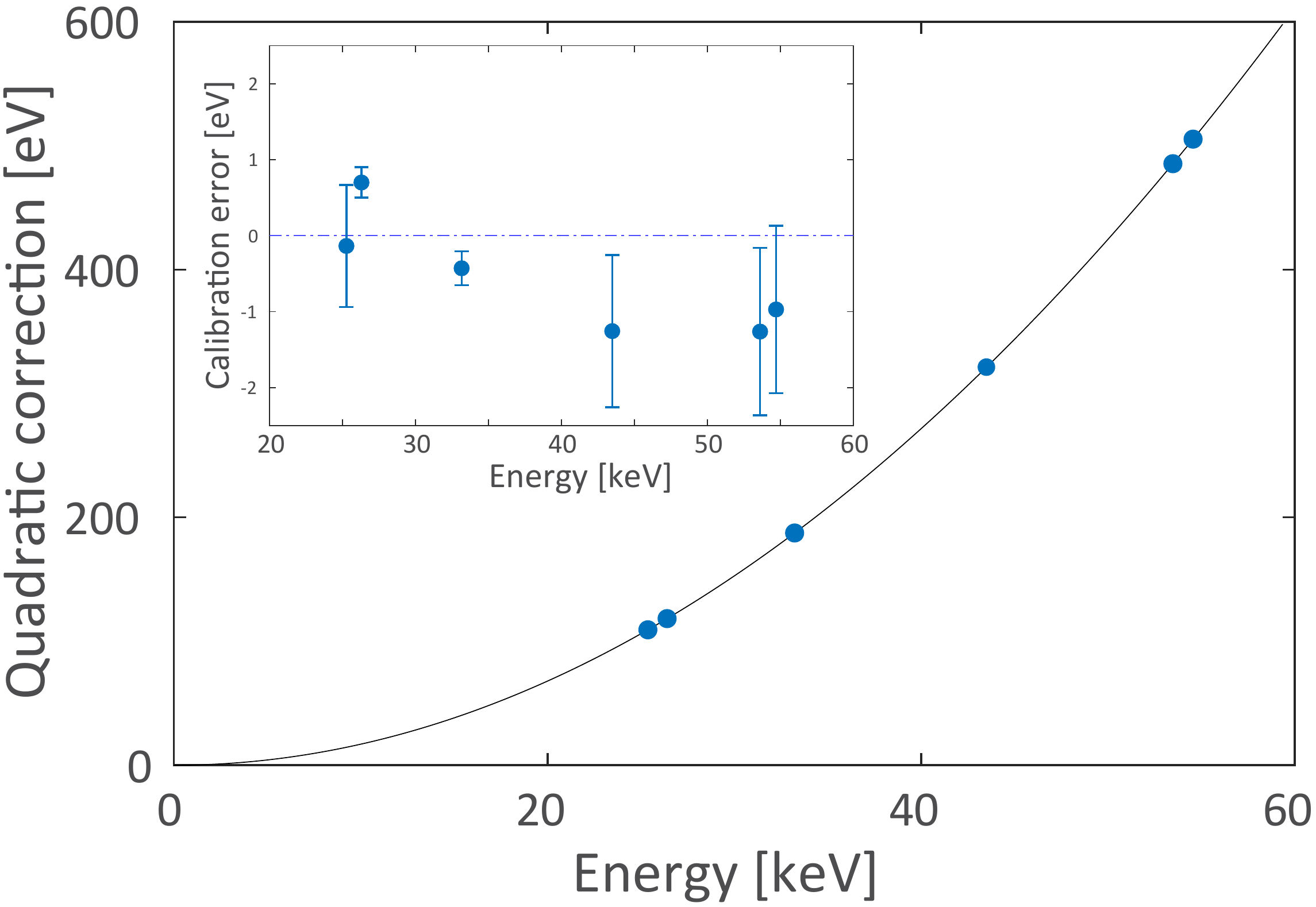}
\caption{Residual uncertainty of the calibration lines with respect to the literature values (see \cref{tab:referencelines}) before the quadratic correction. The residuals follow the quadratic polynomial, indicating that quadratic correction is sufficient. Inset: residual energy differences after the quadratic correction.}
\label{calib}
\end{figure}

The raw amplitude data obtained from each pixel are corrected for temperature, which is extracted from the simultaneously triggered asymmetric pixel-pairs located in each of the four corners of the detector (see supplementary material). The data from individual pixels $(p)$ are corrected quadratically $E(p)=a(p)\times \left(U(p)+b(p)U^2(p) \right )$ to account for small differences in the individual pixel's gain characteristics and combined into a single dataset $E$~\cite{Bates2016}. The \mbox{maXs-30} shows an excellent gain linearity with a nonlinearity of only about $0.6\,\%$ in the energy range of interest, i.e., up to \SI{40}{\kilo\electronvolt} (see \cref{calib}). The nonlinearity precisely follows a second-order polynomial and thus can be perfectly quantified.

To use all the information available in the spectrum and to minimize the free parameters used in the energy calibration, the quadratic terms $b(p)$ are extracted directly from the experimental data. From energy conservation within the nuclear level structure, we identified two decay loops: \SI{54.7}{\kilo\electronvolt}=\SI{25.3}{\kilo\electronvolt}+\SI{29.4}{\kilo\electronvolt} and \SI{54.7}{\kilo\electronvolt}+\SI{13.2}{\kilo\electronvolt}=\SI{25.3}{\kilo\electronvolt}+\SI{42.6}{\kilo\electronvolt} (see \cref{FIG. 1.}a). The quadratic terms $b(p)$ are adjusted to self-consistently fulfill the two conditions above (see supplementary material).

To convert from the amplitude $U$ to energy $E$, we use the reference lines listed in Table~\ref{tab:referencelines}. Experimentally, the linear correction terms for every pixel $a(p)$ are determined by minimizing the squares of the residuals, with the calibration points weighted by the fit and the literature uncertainties (see \cref{calib}). For the calibration, we have chosen only well-resolved gamma lines. We excluded the $^{229}$Th lines that are used in further data analysis and \mbox{x ray} lines because their energy and lineshape might be influenced by the chemical environment~\cite{campbell1990}. Three out of four energy calculations schemes used below are insensitive to the energy calibration (see \cref{eq:iso1,eq:iso2,eq:iso3,eq:iso4}). Either the difference of energy levels (\cref{eq:iso1,eq:iso2}) is used to extract the isomer energy, or their lineshape is analyzed \cref{eq:iso3}. However, the absolute energy scheme using \cref{eq:iso4} is sensitive to the energy calibration, and a different choice of calibration lines can lead to a different result.

Experimental imperfections, i.e., nonlinearities in the analog-to-digital conversion or detector chip inhomogeneity, can lead to small oscillations of the residuals of the energy calibration curve. These are too small to be quantified experimentally (see \cref{calib}). We use the standard deviation of calibration lines from their literature values to estimate the uncertainty due to the local calibrations. The calibration uncertainty of every peak is (\SI{0.76}{\electronvolt}).

\begin{table}[h]
	\centering
	\setlength{\tabcolsep}{1pt}
    \renewcommand{\arraystretch}{1.2}
		\begin{tabular}{l c c c c}
		\hline
		Decay path & Measured\,[eV] & Reference\,[eV] & resid.\,[$\sigma$] & ref.\\
		\hline
		\setlength{\thickmuskip}{0mu}$^{229}$Th$(9/2^+\shortrightarrow7/2^+)$ & 25314.4(8) & 25314.6(8) & 0.2 &\cite{Helmer1994} \\
		\setlength{\thickmuskip}{0mu}$^{237}$Np$(5/2^-\shortrightarrow7/2^+)$ & 26345.3(8) & 26344.6(2) & 3.4 & \cite{Helmer2000} \\
		\setlength{\thickmuskip}{0mu}$^{237}$Np$(7/2^+\shortrightarrow5/2^+)$ & 33195.8(8) & 33196.3(2) & 2.2 & \cite{Basunia2006} \\
		\setlength{\thickmuskip}{0mu}$^{234}$U $(2^+\rightarrow0^+)$ & 43496.8(8) & 43498.1(10) & 1.3 & \cite{browne2007nuclear} \\		\setlength{\thickmuskip}{0mu}$^{229}$Th$(9/2^+\shortrightarrow7/2^+)$ & 53609.3(8) & 53610.7(11) & 1.2 & \cite{Helmer1994} \\
		\setlength{\thickmuskip}{0mu}$^{229}$Th$(9/2^+\shortrightarrow7/2^+)$ & 54703.0(8) & 54704.0(11) & 0.9 & \cite{Barci2003} \\
		\hline
		\end{tabular}
	\caption{Reference lines used in the calibration. The uncertainty for measured values is dominated by the calibration uncertainity (\SI{0.76}{\electronvolt}). The contribution of the statistical uncertainty of the fit is negligible. The deviation from the literature values is reported as the number of standard deviations from the reference values [$\sigma$].}
	\label{tab:referencelines}
\end{table}

While higher-order nonlinearities contribute to the uncertainty of each peak's absolute energy, there is one more artefact affecting the lineshapes that we have to consider during data analysis. The electronic signal after the photon absorption decays faster than the temperature of the pixel. This is caused by a differential readout of the pixel-pairs and design details of the pair-wise heat-sinking of pixels~\cite{Fleischmann2009,Kozorezov2013}. If a new event occurs in the pixel-pair before it cooled down to idle state, the signal response to energy input is reduced, leading to a reduced signal height, and low energy tails in the spectrum. We mitigated this effect during the experiment by applying the hold-off mentioned above, but small distortions from a Gaussian lineshape are still present.

To adequately describe the data, we use a generic lineshape based on an asymmetric Voigt profile (see supplementary material) that can fit all the gamma lines in the spectrum. The Gaussian function, characterized by variance ($\sigma)$, is convolved with a wider Lorentzian profile on the low energy side $(\gamma_1)$ and a narrower Lorentzian profile on the high energy side $(\gamma_2)$. All three parameters follow a weak quadratic dependence on the energy of the line $\sigma=\sigma^0\left ( 1+cE^2 \right )$ and $\gamma_{1(2)}=\gamma_{1(2)}^0\left ( 1+cE^2 \right )$. These four parameters $(\sigma, \gamma_1, \gamma_2, c)$ were extracted from a simultaneous fit of all $\gamma$ lines; they are common to all the $\gamma$ lines in the spectrum. For each individual line fit, only two free parameters are varied, the amplitude ($A$) and the energy ($E$).
\begin{figure}[h!]
\includegraphics[width=\linewidth]{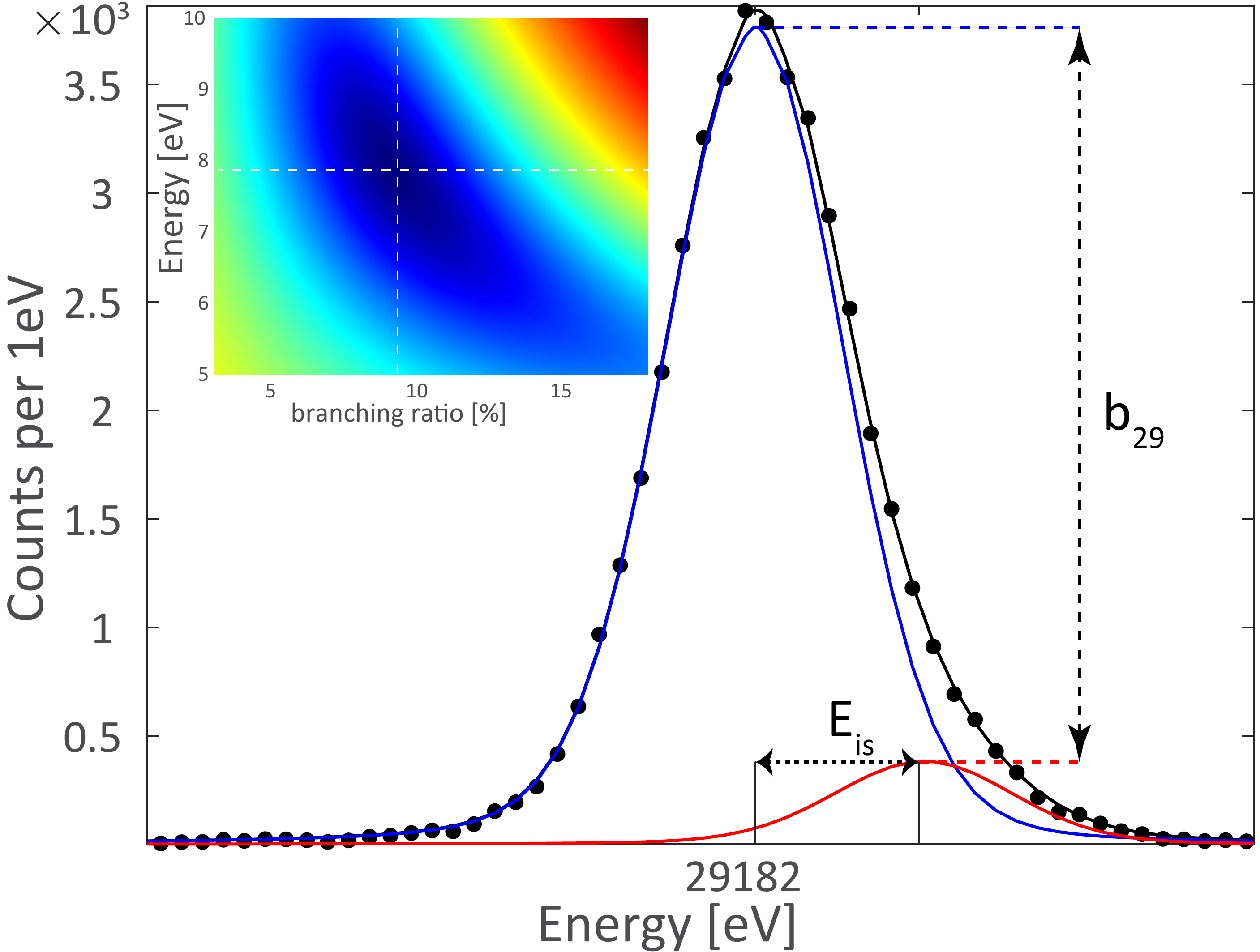}
\caption{Lineshape of the \SI{29.2}{\kilo\electronvolt} doublet. The red curve represents the fit of the \SI{29190}{\electronvolt} line, and the blue curve represents the fit of the \SI{29182}{\electronvolt} line. The branching ratio $b_{29}$ and the isomer energy ($E_{\rm is}$) can be extracted directly from this fit. The inset shows a 2-D plot of $\chi^2$ as a function of branching ratio and the isomer energy. The white dashed lines point to the fitted $E_{\rm is}$ and $b_{29}$ values.}
\label{lineshape}
\end{figure}

\begin{figure}
\includegraphics[width=\linewidth]{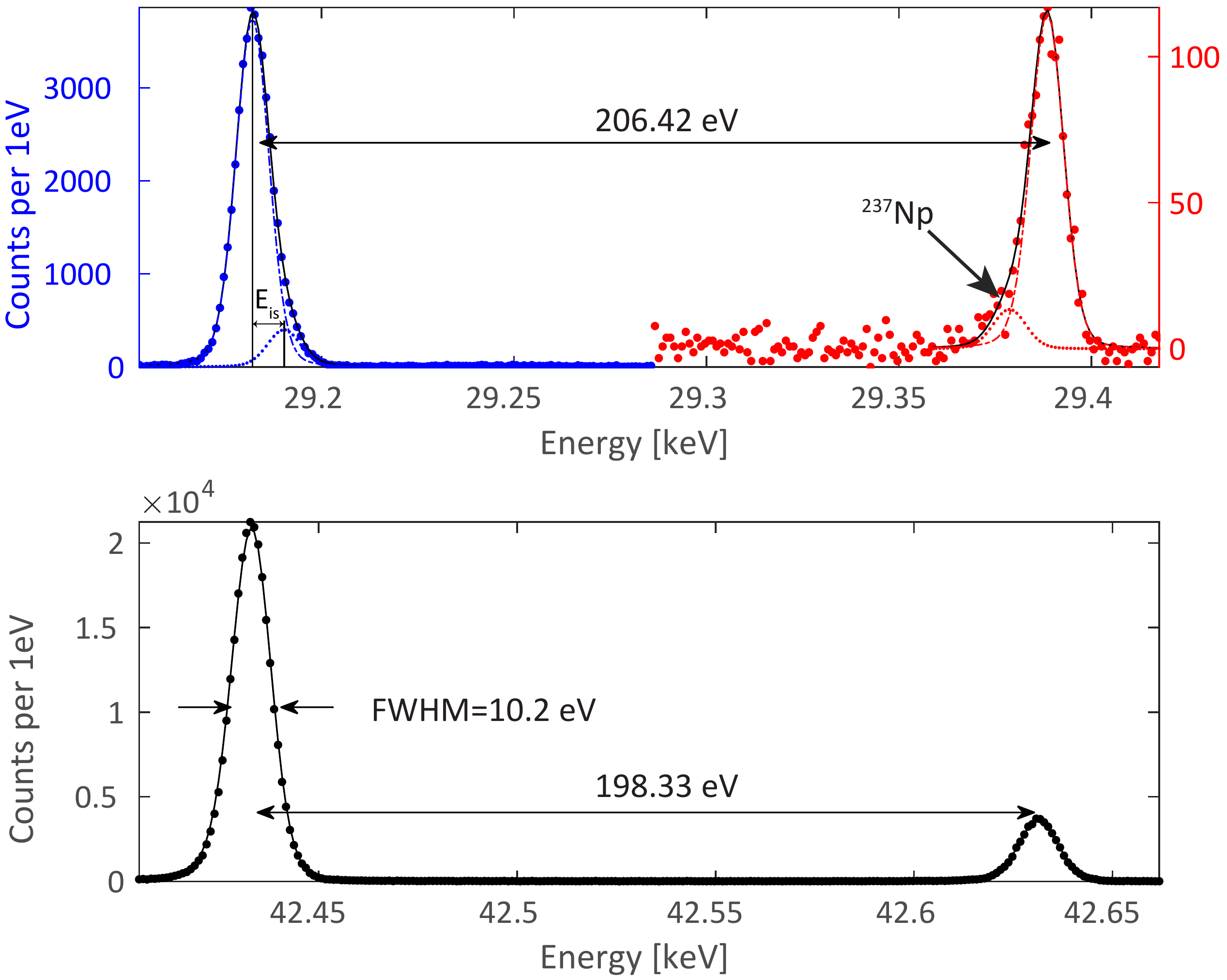}
\caption{MMC spectra of the \SI{29}{\kilo\electronvolt} and \SI{42}{\kilo\electronvolt} doublets. The $^{237}$Np contamination leads to a weak signal $\left(\SI{29378.6\pm1.8}{\electronvolt}\right)$ which overlaps the \SI{29.4}{\kilo\electronvolt} line (see supplementary material).}
\label{differences}
\end{figure}

The energy resolution of $\sim$\SI{10}{\electronvolt} (see \cref{FIG. 1.}a) together with the outstanding linearity (see \cref{calib}) of the \mbox{maXs-30} micro-calorimeter allows us to perform four different types of data analysis to extract the $^{229}$Th isomer energy. With high resolution, we can partly resolve the \SI{29.2}{\kilo\electronvolt} doublet line. By analyzing its lineshape deviations compared to isolated lines in the spectrum, we can extract both the isomer energy ($E_{\mathrm{is}}$) and the branching ratio ($b_{29}$). To analyze the line doublets, we use a pair of generic lineshapes from \cref{differences} and let the doublet splitting and their relative amplitudes as free parameters. From their relative amplitudes, we measure that the $5/2^+$ state has a significant inter-band decay probability of $b_{29}=9.3(6)\%$, which leads to a doublet shape of the \SI{29.2}{\kilo\electronvolt} line with a splitting equal to the isomer energy (see \cref{FIG. 1.}a). We find a value of,
\begin{equation}
E_{\rm is,lineshape}=\SI{7.84\pm0.29}{\electronvolt}.
\label{eq:iso3}
\end{equation}

The \SI{42.4}{\kilo\electronvolt} line shows a very small lineshape deviation from a monoenergetic line. This allowed us to put an upper one-sigma bound on the branching ratio $b\left ( \frac{42.4\rightarrow 0.08\ \mathrm{keV}}{42.4\rightarrow 0\ \mathrm{keV}} \right )=b_{42}<0.7\%$. These branching ratios are consistent with the previous experimental value for $b_{29}=9.3(6)\%$~\cite{Masuda2019}, and $b_{42}$ lies in the range of theoretical predictions $0.2\%<b_{42}<2\%$~\cite{Tkalya2015}. There is a strong correlation between the $E_{\rm is,lineshape}$ and $b_{29}$. A more accurate independent measurement of $b_{29}$ combined with our experiment would further decrease the uncertainty of the isomer energy.

The second analysis makes use of the two pairs of closely spaced lines at \SI{29}{\kilo\electronvolt} and \SI{42}{\kilo\electronvolt}. Measuring the distances of these two pairs yields the isomeric state energy $(E_{\rm is})$ (see \cref{differences}). A previous experiment performed with a silicon semiconductor micro-calorimeter extracted $E_{\rm is}$ using this scheme, however, with a much higher uncertainty due to the limited resolution and higher nonlinearities~\cite{Beck2007,Beck09}.

The absolute energy scheme uncertainty of each peak is dominated by the calibration uncertainty (\SI{0.76}{\electronvolt}). Because the calibration uncertainty is a slowly varying function of the energy, it is significantly compensated when subtracting energies of closely spaced lines $\Delta$E$_{29}$ and $\Delta$E$_{42}$ (see \cref{differences}). The lines \SI{29.18}{\kilo\electronvolt} and \SI{42.43}{\kilo\electronvolt}, due to their doublet nature, are fitted with two generic lineshapes each. The relative amplitudes of these functions are set according to known inter-band branching ratios and the spacing is $E_{\rm is}$ (see supplementary material). The inter-band branching ratios are $b_{29}=9.3(6)\%=\frac{1}{9.8(6)}$ and $b_{42}=0.3(3)\%=\frac{1}{305}$~\cite{Masuda2019,Beck09}.

We obtain the isomer energy as,
\begin{equation}
E_{\rm is,1} = \SI{206.42\pm0.17}{\electronvolt}-\SI{198.33\pm0.02}{\electronvolt}=\SI{8.10\pm0.17}{\electronvolt}.
\label{eq:iso1}
\end{equation}
The uncertainty emerges from the statistics of the weak \SI{29.4}{\kilo\electronvolt} line (\SI{0.12}{\electronvolt}), from the uncertainty of the branching ratio (\SI{0.03}{\electronvolt}), and from the uncertainty of the $^{237}$Np contamination (\SI{0.12}{\electronvolt}), see supplementary material.

\begin{figure}
\includegraphics[width=\linewidth]{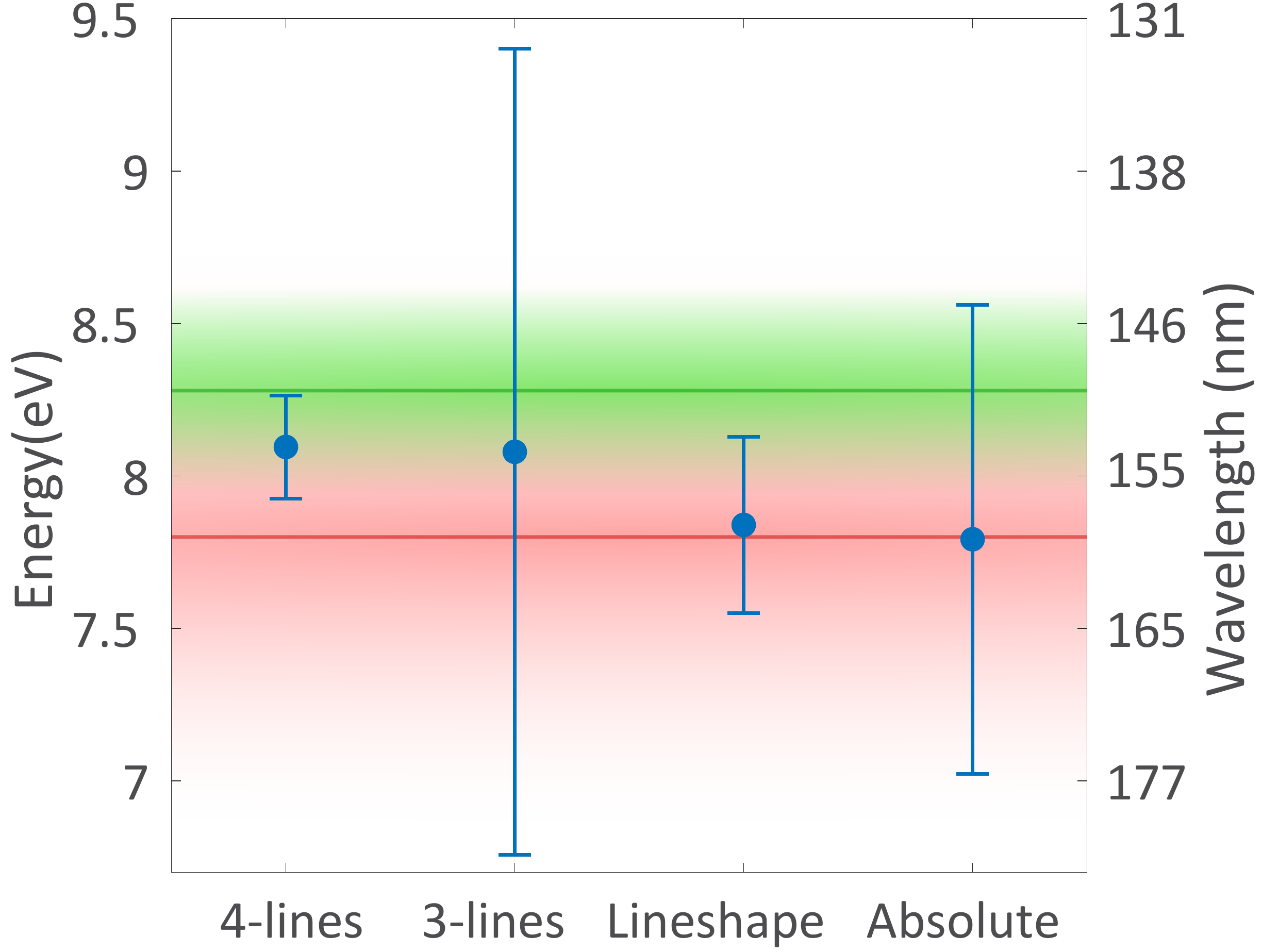}
\caption{Isomer energies $E_{\rm is}$ measured in this study compared to previous experiments using $\gamma$ spectroscopy. The green and red faded areas represent the isomer energy reported in~\cite{Beck09,Seiferle2019}, respectively, with their corresponding uncertainties. Error bars for our $E_{\rm is}$ represent the root sum square of statistical and systematic uncertainty.}
\label{FIG. 5.}
\end{figure}

Alternatively we can extract the isomer energy as,
\begin{equation}
\resizebox{0.89\hsize}{!}{%
$E_{\rm is,2}=\SI{42433.1}{\electronvolt}-\SI{13242.9}{\electronvolt}-\SI{29182.1}{\electronvolt}=\SI{8.1\pm1.3}{\electronvolt}$}.%
\label{eq:iso2}
\end{equation}
This method has the advantage of avoiding the weak inter-band \SI{29.4}{\kilo\electronvolt} transition (see \cref{FIG. 1.}a). The disadvantage is that there are no closely spaced line pairs. Therefore, the calibration uncertainty is not compensated but adds up. This result is consistent with the analysis of~\cref{eq:iso1}, but the uncertainty is much higher (see \cref{FIG. 5.}).

In a recent experiment, synchrotron radiation was used to excite the ground state of $^{229}$Th to the (5/2$^+$) state (inter-band transition), the excitation energy was measured as \SI{29189.93\pm0.07}{\electronvolt}~\cite{Masuda2019}. In this work, we have accurately measured the intra-band transition from the (5/2$^+$) state to the isomer state: \SI{29182.1\pm0.8}{\electronvolt} (see \cref{FIG. 1.}). Subtracting these two values yields a fourth value for E$_{\rm is}$,
\begin{equation}
\resizebox{0.89\hsize}{!}{%
	$E_{\rm is,abs} = \SI{29189.93\pm0.07}{\electronvolt}-\SI{29182.1\pm0.8}{\electronvolt}=\SI{7.8\pm0.8}{\electronvolt}$}.%
\label{eq:iso4}
\end{equation}

The estimated uncertainty is dominated by the calibration uncertainty (\SI{0.76}{\electronvolt}). This scheme was recently used in another experiment, reporting an isomer energy of ($E_{\rm is}=\SI{8.30\pm0.92}{\electronvolt}$)~\cite{Yamaguchi2019}.

In conclusion, the energy of the $^{229}$Th isomer state was measured by recording a high resolution (FWHM$\,\approx$\,\SI{10}{\electronvolt}) high bandwidth ($\sim$\SI{60}{\kilo\electronvolt}) $\gamma$ spectrum using a cryogenic magnetic micro-calorimeter. We extracted the isomer energy using four different schemes. A comparison of all four results with previous experiments is summarized in \cref{FIG. 5.}. Weighting these results with their statistical and systematic uncertainty and combining them, we constrain the one-sigma interval for the isomer energy to be \SI{7.88}{\electronvolt}$<E_{\rm is}<$\SI{8.16}{\electronvolt}. We also measured the branching ratio of the second excited state $b_{29}=9.3(6)\%$ and found it to be compatible with previously measured values.
\begin{acknowledgments}
This work was supported by the European Union's Horizon 2020 research and innovation program under grant agreement No. 664732 "nuClock", grant agreement No. 856415 "ThoriumNuclearClock", grant agreement No. 882708 "CrystalClock" and grant agreement No. 824109 "European Microkelvin Platform". The project has also received funding from the EMPIR program co-financed by the Participating States and from the European Union's Horizon 2020 research and innovation program. We thank N. Trautmann for his contributions to the source preparation and the staff of the mechanical workshop of the Institute of Nuclear Chemistry in Mainz for the construction of the $^{233}$U liquid-source container, J. Schwestka for sample preparation at TU Wien and S. Stellmer and P. Thirolf for helpful discussions.

\end{acknowledgments}
\bibliography{ms.bib}
\end{document}


\title{Measurement of the \texorpdfstring{$^{229}$}{229}Th isomer energy with a magnetic micro-calorimeter}
\author{Tomas Sikorsky}
\thanks{T.S. and J.G. contributed equally to this work.}
\affiliation{Kirchhoff-Institute for Physics, Heidelberg University, INF 227, 69120 Heidelberg, Germany}
\affiliation{Institute for Atomic and Subatomic Physics, TU Wien, Stadionallee 2, 1020, Vienna, Austria}
\author{Jeschua Geist}
\thanks{T.S. and J.G. contributed equally to this work.}
\author{Daniel Hengstler}
\author{Sebastian Kempf}
\author{Loredana Gastaldo}
\author{Christian Enss}
\affiliation{Kirchhoff-Institute for Physics, Heidelberg University, INF 227, 69120 Heidelberg, Germany}
\author{Christoph Mokry}
\affiliation{Johannes Gutenberg University, 55099 Mainz, Germany.}
\affiliation{Helmholtz Institute Mainz, 55099 Mainz, Germany.}
\author{J\"org Runke}
\affiliation{Johannes Gutenberg University, 55099 Mainz, Germany.}
\affiliation{GSI Helmholtzzentrum f\"ur Schwerionenforschung GmbH, 64291 Darmstadt, Germany}
\author{Christoph E. D\"ullmann}
\affiliation{Johannes Gutenberg University, 55099 Mainz, Germany.}
\affiliation{Helmholtz Institute Mainz, 55099 Mainz, Germany.}
\affiliation{GSI Helmholtzzentrum f\"ur Schwerionenforschung GmbH, 64291 Darmstadt, Germany}
\author{Peter Wobrauschek}
\author{Kjeld Beeks}
\author{Veronika Rosecker}
\author{Johannes H. Sterba}
\author{Georgy Kazakov}
\author{Thorsten Schumm}
\affiliation{Institute for Atomic and Subatomic Physics, TU Wien, Stadionallee 2, 1020, Vienna, Austria}
\author{Andreas Fleischmann}
\affiliation{Kirchhoff-Institute for Physics, Heidelberg University, INF 227, 69120 Heidelberg, Germany}
\date{\today}
\maketitle
\begin{center}
	\textbf{Sample contamination}
\end{center}
The $^{233}$U source was chemically purified to remove daughter products of $^{241}$Am and $^{233}$U. After purification, we detected molar ratios $\sfrac{^{232}\rm U}{^{233}\rm U}=2\times10^{-7}$, $\sfrac{^{234}\rm U}{^{233}\rm U}=1\times10^{-3}$, $\sfrac{^{238}\rm Pu}{^{233}\rm U}=7\times10^{-6}$, $\sfrac{^{240}\rm Pu}{^{233}\rm U}=1\times10^{-3}$, $\sfrac{^{241}\rm Am}{^{233}\rm U}=7\times10^{-4}$ and $\sfrac{^{237}\rm Np}{^{233}\rm U}=$\num{8.7(17)e-4}. The $^{237}$Np contamination is important because it produces a line at $\SI{29378.6}{\electronvolt}$, which overlaps with the \SI{29388.5}{\electronvolt} line from the $^{233}$U decay.
\begin{figure}[ht]
\includegraphics[width=\linewidth]{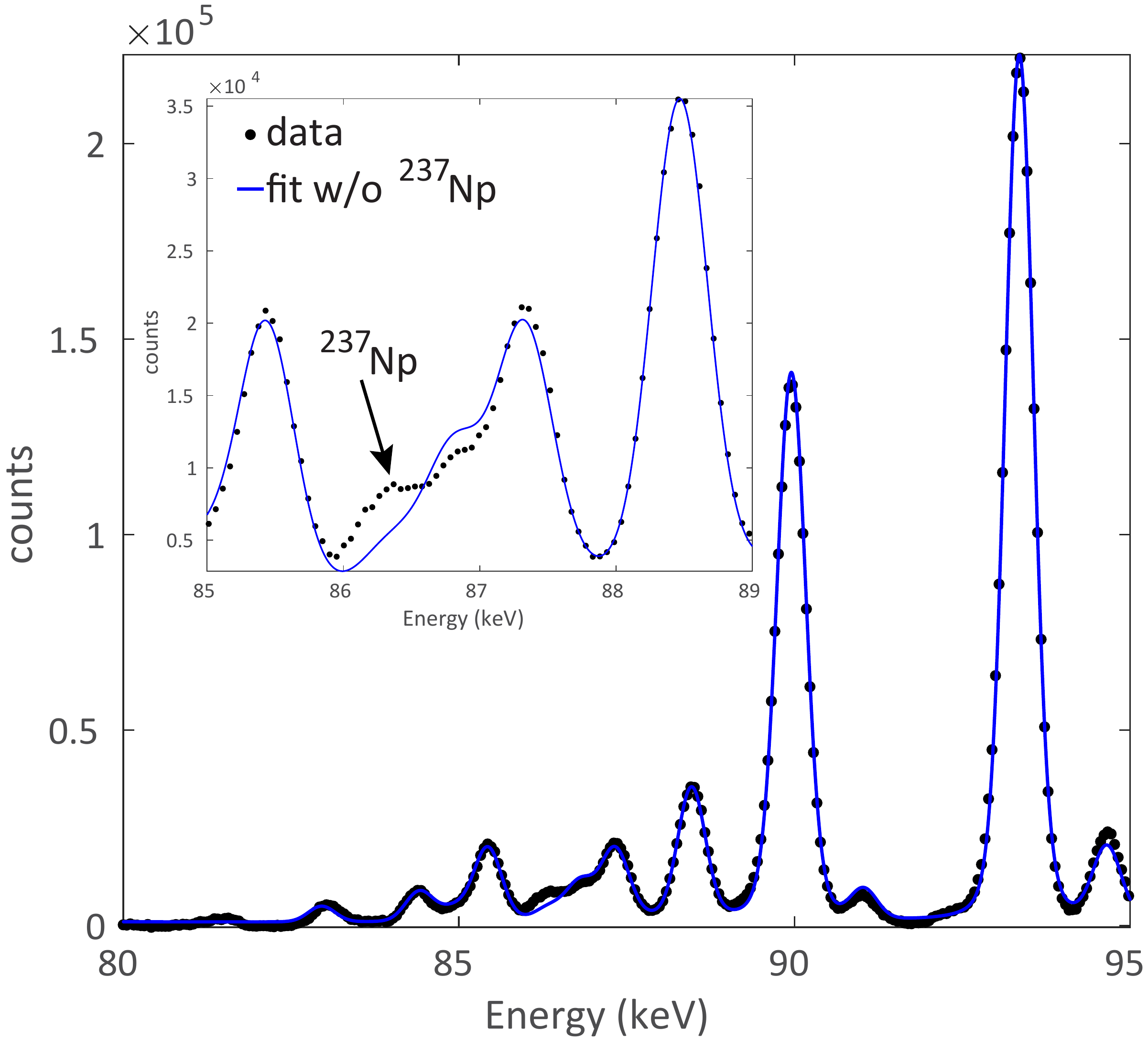}
\caption{A $\gamma$ spectrum of the sample recorded using an HPGe detector. After subtracting $^{232,233,234}$U and their daughter elements, a signal at \SI{86.5}{\kilo\electronvolt} remains.}
\label{neptunium}
\end{figure}

This appears in the spectrum as an anomalous low energy tail of the \SI{29.4}{\kilo\electronvolt} line. To compensate for the $^{237}$Np background, we need to measure its concentration. To get the concentration we fit the \SI{29.4}{\kilo\electronvolt} line with a doublet of generic lineshape with relative amplitude and position as free parameters. We extract the concentration of $\sfrac{^{237}\rm Np}{^{233}\rm U}=$\num{8.8(19)e-4}. To confirm the ${^{237}\rm Np}$ concentration we recorded another $\gamma$ spectrum in the energy range 0--\SI{100}{\kilo\electronvolt} using a liquid nitrogen-cooled High Purity Ge detector (HPGe) Intertechnique (see \cref{neptunium}). In the spectrum we identify the $^{233}$Pa$(5/2^+\rightarrow3/2^-)$ \SI{86.5}{\kilo\electronvolt} line. From the intensity of the line we obtain the molar ratio of $\sfrac{^{237}\rm Np}{^{233}\rm U}=$\num{8.7(37)e-4}. Combining these two measurements we obtained the $\sfrac{^{237}\rm Np}{^{233}\rm U}=$\num{8.7(17)e-4} molar concentration, which was used in the data analysis.

\begin{center}
	\textbf{Signal processing}
\end{center}
The signal rise after the $\gamma$ photon absorption event is determined by the thermal conductivity of the region between the absorber and the sensor, resulting in an expected signal rise time of about $\tau=\SI{9.6}{\micro\second}$. The AC signal decay is \SI{0.75}{\milli\second} and DC signal decays in \SI{2.7}{\milli\second}.

\begin{figure}[ht]
\includegraphics[width=\linewidth]{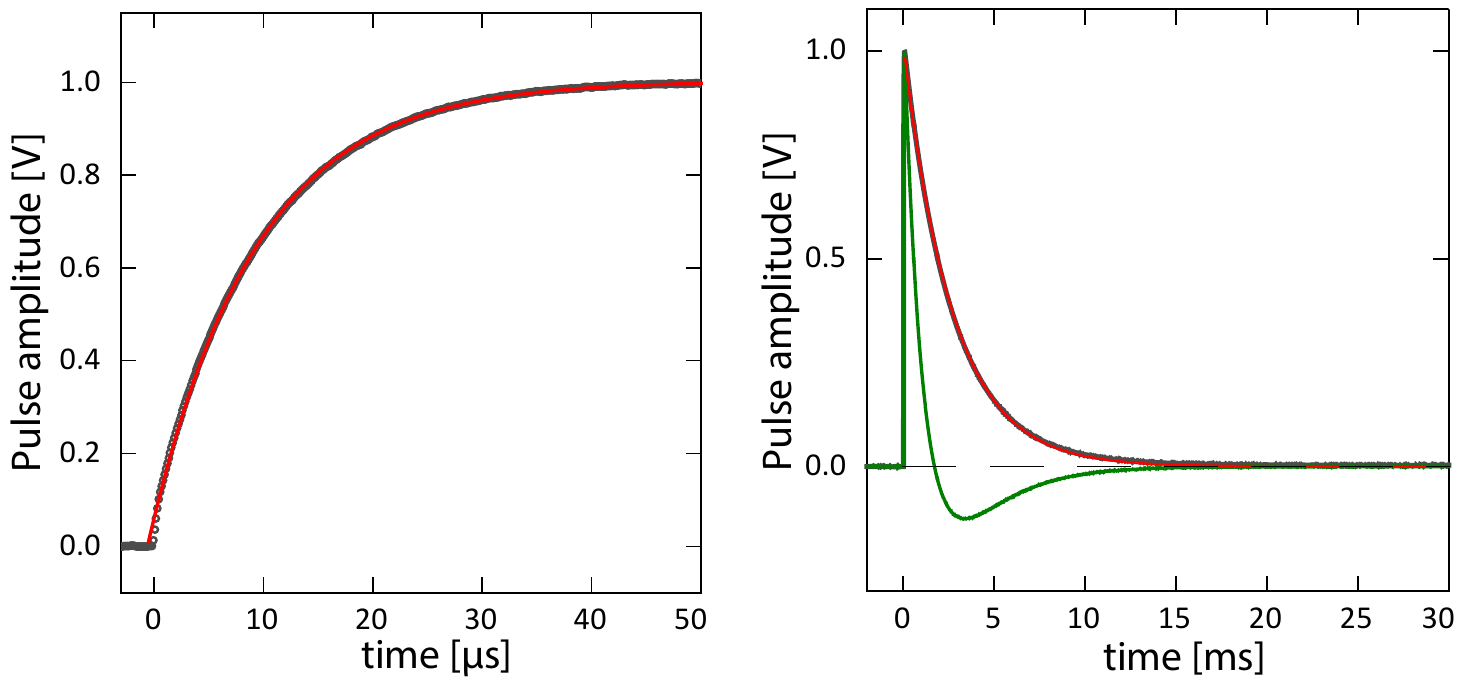}
\caption{Left: Closeup of the first \SI{50}{\micro\second} of a single pulse after the absorption of a $\gamma$ photon. The red line shows the exponential fit $\tau=\SI{9.6}{\micro\second}$. Right: The first \SI{30}{\milli\second} of the signal decay. The red line shows the exponential fit of the DC-coupled signal, and the green line shows the AC-coupled signal.}
\label{rise_fall}
\end{figure}

\begin{center}
	\textbf{Detector anatomy}
\end{center}
The maXs-30 detector was fabricated at Kirchhoff Institute for Physics, Heidelberg~\cite{phdthesis}. It is cooled to an operating temperature of 12\,mK by a dry dilution refrigerator. The $^{233}$U source is placed outside the cryostat behind a polyimide window, yielding a count rate of $\approx 2$\,Hz per pixel after hold-off. The detector-signal is amplified by 16-SQUID-series-array mounted at the 4 Kelvin stage of the cryostat. These SQUIDs have also been produced in-house~\cite{Kempf2015}. The two-stage SQUID channels are read out at room temperature by SQUID-electronics of type XXF-1 and a 16-channel digitizer card SIS3316 with a bandwidth of up to 125\,MHz and a resolution of 16\,bit.
\begin{figure}[ht]
\includegraphics[width=\linewidth]{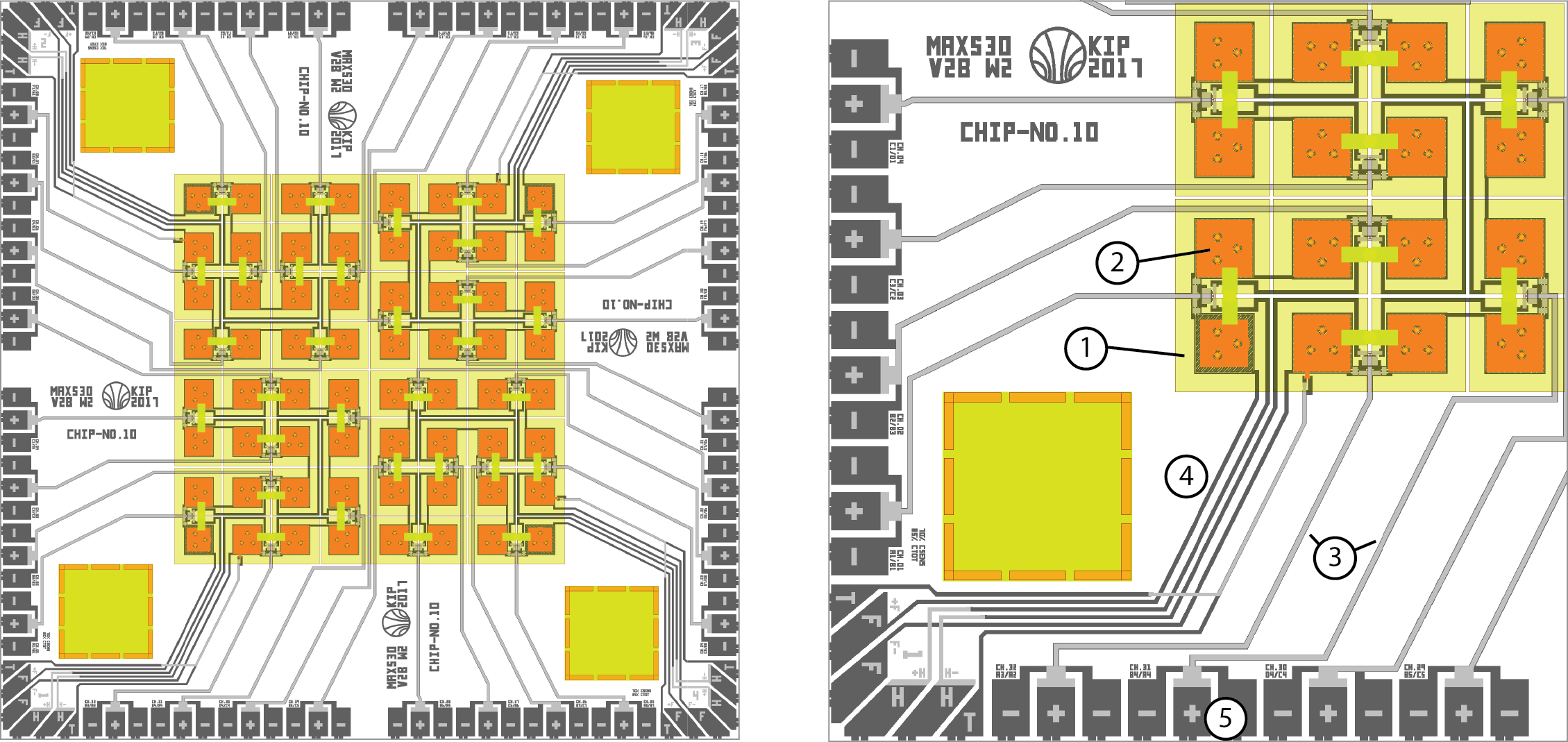}
\caption{Layout of the maXs-30 detector chip with a closeup showing one quadrant of the chip: (1) absorber; (2) temperature sensor; (3) read-out
lines; (4) connecting lines for the persistent current; (5) contacts for the individual lines.}
\label{drawing}
\end{figure}
It is composed of 8\,$\times$\,8 pixels with absorbers made of gold providing a total active area of \SI{4}{\milli\meter}$\times$\SI{4}{\milli\meter}. The pixels are grouped in pairs where the difference of the temperature of two detector pixels is read out (see \cref{drawing}). The absorber thickness has been adjusted to \SI{20}{\micro\meter}, resulting in a 65\,\% stopping power at \SI{30}{\kilo\electronvolt} to reach a good compromise between energy resolution and detection efficiency. In each of the four corners, a pixel-pair allows to detect and correct for temperature fluctuations of the detector chip.

\begin{figure}[ht]
\includegraphics[width=\linewidth]{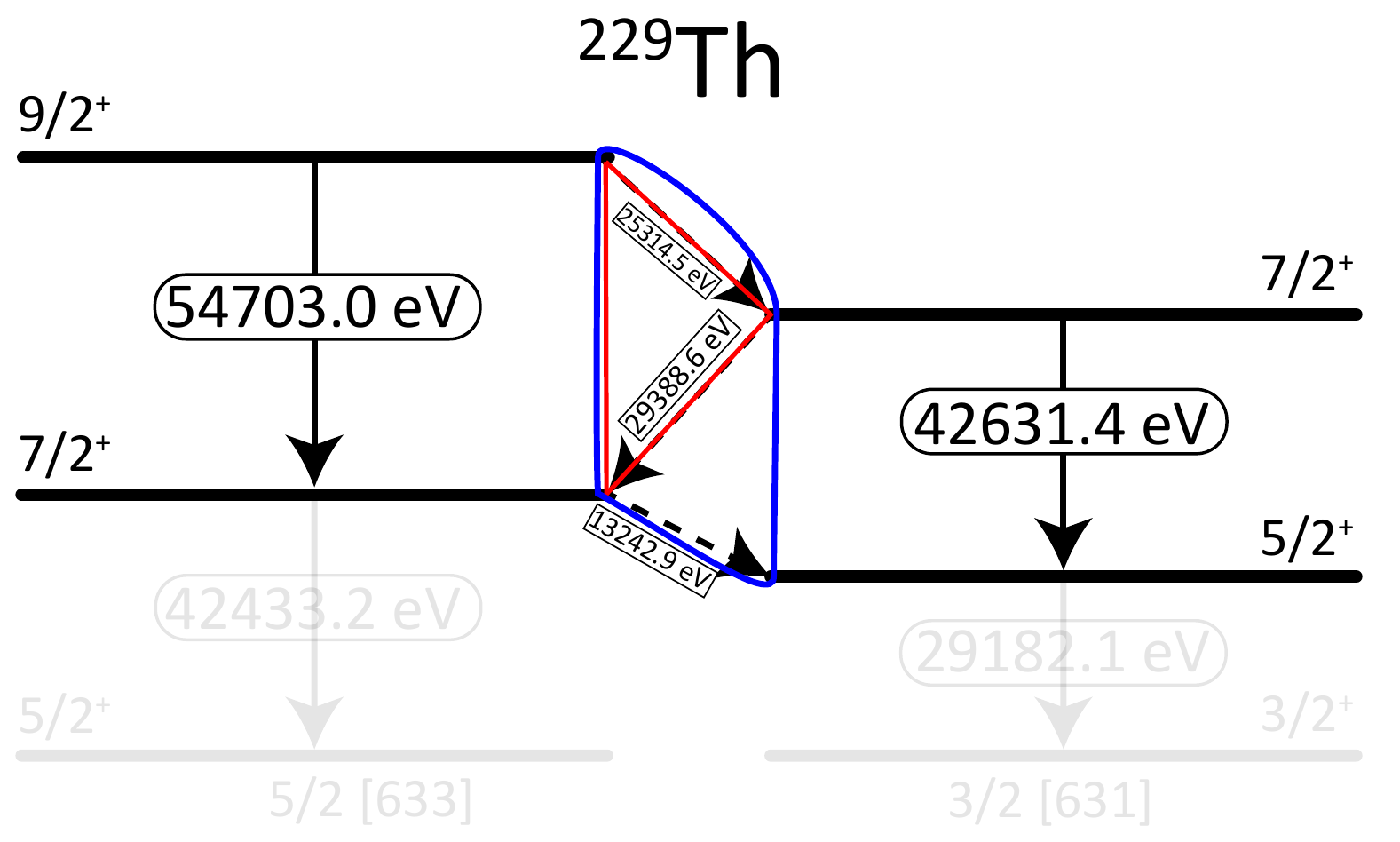}
\caption{Partial nuclear level scheme of the $^{229}$Th with relevant decay paths and energies. The levels that were used for the calibration of the quadratic term $(b)$ are highlighted.}
\label{calibration}
\end{figure}

\begin{figure}[ht]
\includegraphics[width=\linewidth]{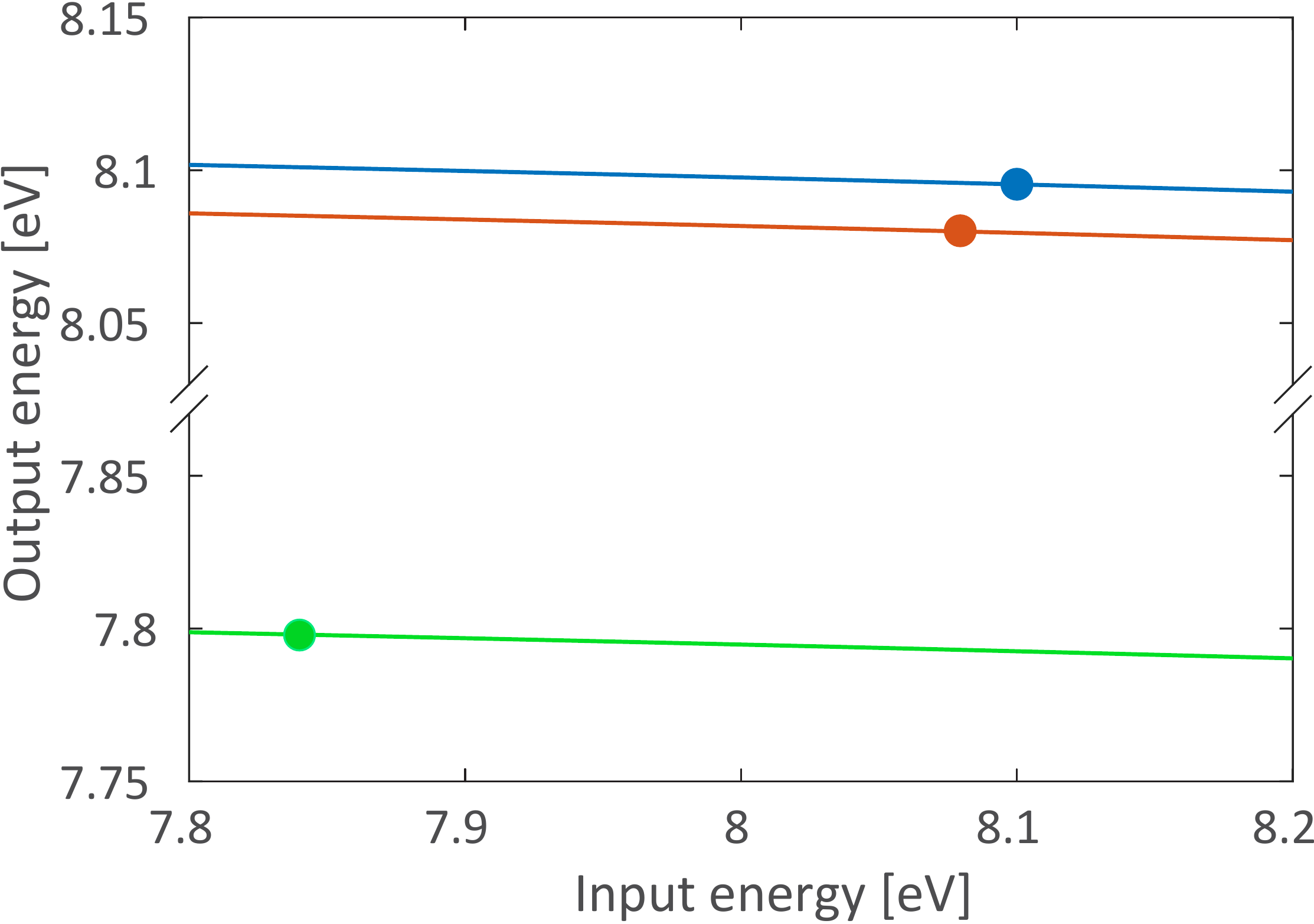}
\caption{The sensitivity of the output isomer energy ($E_{\mathrm{is}}$) on the input isomer energy. The blue line represents the 4-lines scheme, the red line represents the 3-lines scheme, and the green line represents the absolute energy calculation scheme.}
\label{sensitivity}
\end{figure}

Since the $^{233}$U source cannot be installed inside the cryostat, the outer shield has been equipped with an x ray window made of 150-\si{\micro\meter} thick polyimide, enabling high-energy photons to enter the cryostat. The inner radiation shields have holes with x ray windows made of 6-\si{\micro\meter} thin mylar foil coated with about \SI{40}{\nano\meter} aluminum. A 1-\si{\milli\meter} thick aluminum container encloses the detector platform and acts as a superconducting shield against external electromagnetic perturbations. Additionally, it reduces the count rate for photons below the $\gamma_{2,1}$ line at 29.19\,keV where the spectrum is dominated by x rays from electron shell transitions.
\begin{center}
	\textbf{Energy calibration}
\end{center}
The energy calibration was constrained to maintain the energy conservation in the $^{229}$Th nuclear level structure. The five transitions that were used are shown in \cref{calibration}.
\begin{center}
	\textbf{Self-consistent isomer energy calculation}
\end{center}
The absolute, the 3-lines, and the 4-lines schemes require an isomer energy as an input. For each scheme, we employ the self-consistent iterative approach to calculate the energy. This method is very robust, as the sensitivity of the output energy on the input energy is quite low (\SI{0.02}{\electronvolt\per\electronvolt}) (see \cref{sensitivity}).
\begin{center}
	\textbf{Generic lineshape}
\end{center}
\begin{figure}[ht]
\includegraphics[width=\linewidth]{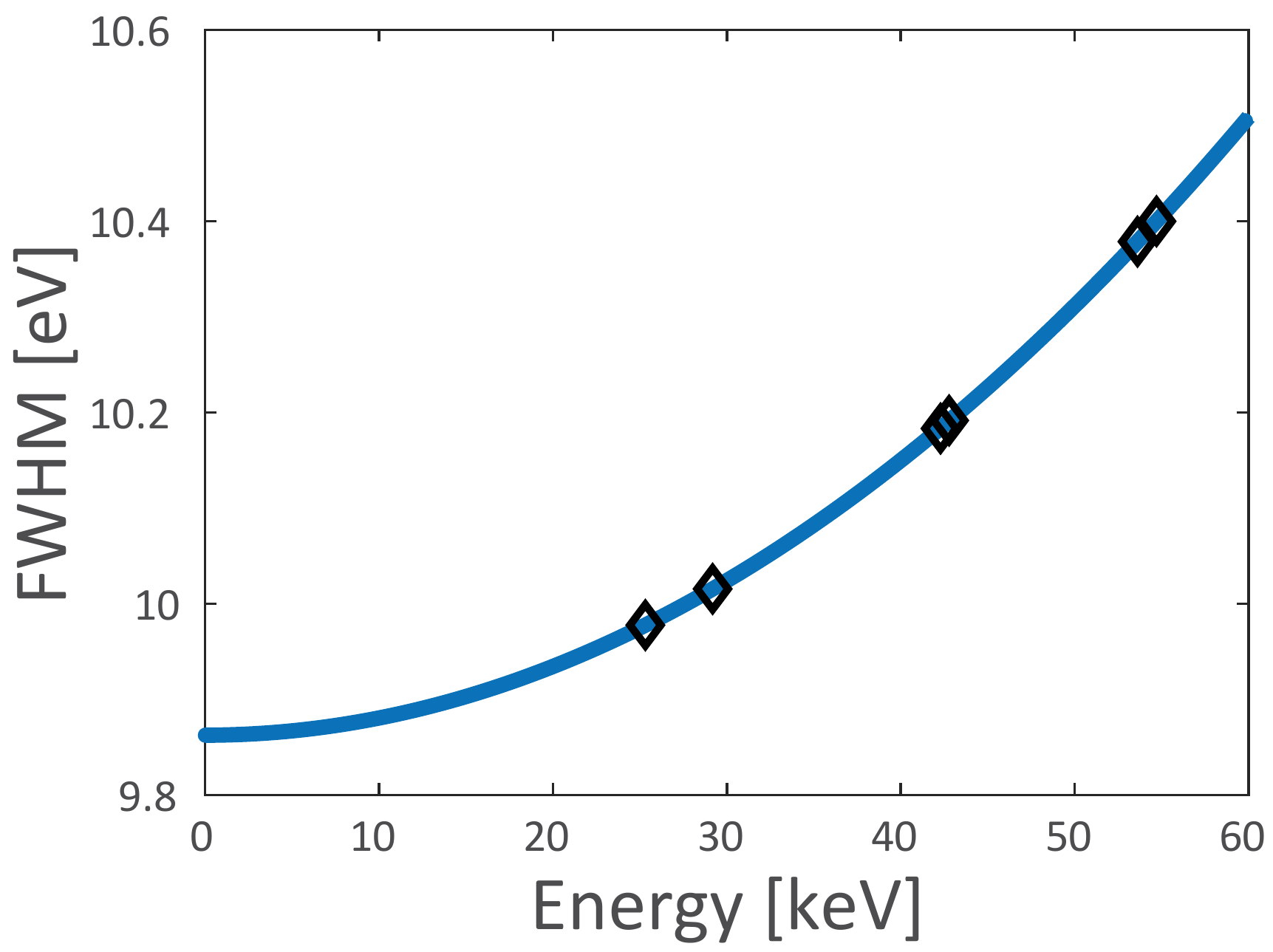}
\caption{Full width at half maximum (FWHM) of the generic lineshape as a function of energy. The linewidth scales with energy as FWHM=\SI{9.87}{\electronvolt}$+c\cdot{E^2}.$ At the \SI{29.2}{\kilo\electronvolt} the FWHM is \SI{10.0}{\electronvolt}}
\label{fwhm}
\end{figure}
The asymmetric Voigt profile is constructed by asymmetrically convoluting a Lorentz with a Gaussian distribution (see \cref{voigt}). The Gaussian distribution $(\sigma^0=\SI{4.06}{\electronvolt})$ is convoluted with a Lorentz profile $(\Gamma_1^0=\SI{0.72}{\electronvolt})$ on the low energy side and $(\Gamma_2^0=\SI{0.52}{\electronvolt})$ on the high energy side. These terms scale quadratically with energy as $c=\SI{1.8E-10}{\per\electronvolt\squared}$ (see \cref{fwhm}).
\begin{widetext}
\begin{center}
\begin{figure}
\includegraphics[width=17cm]{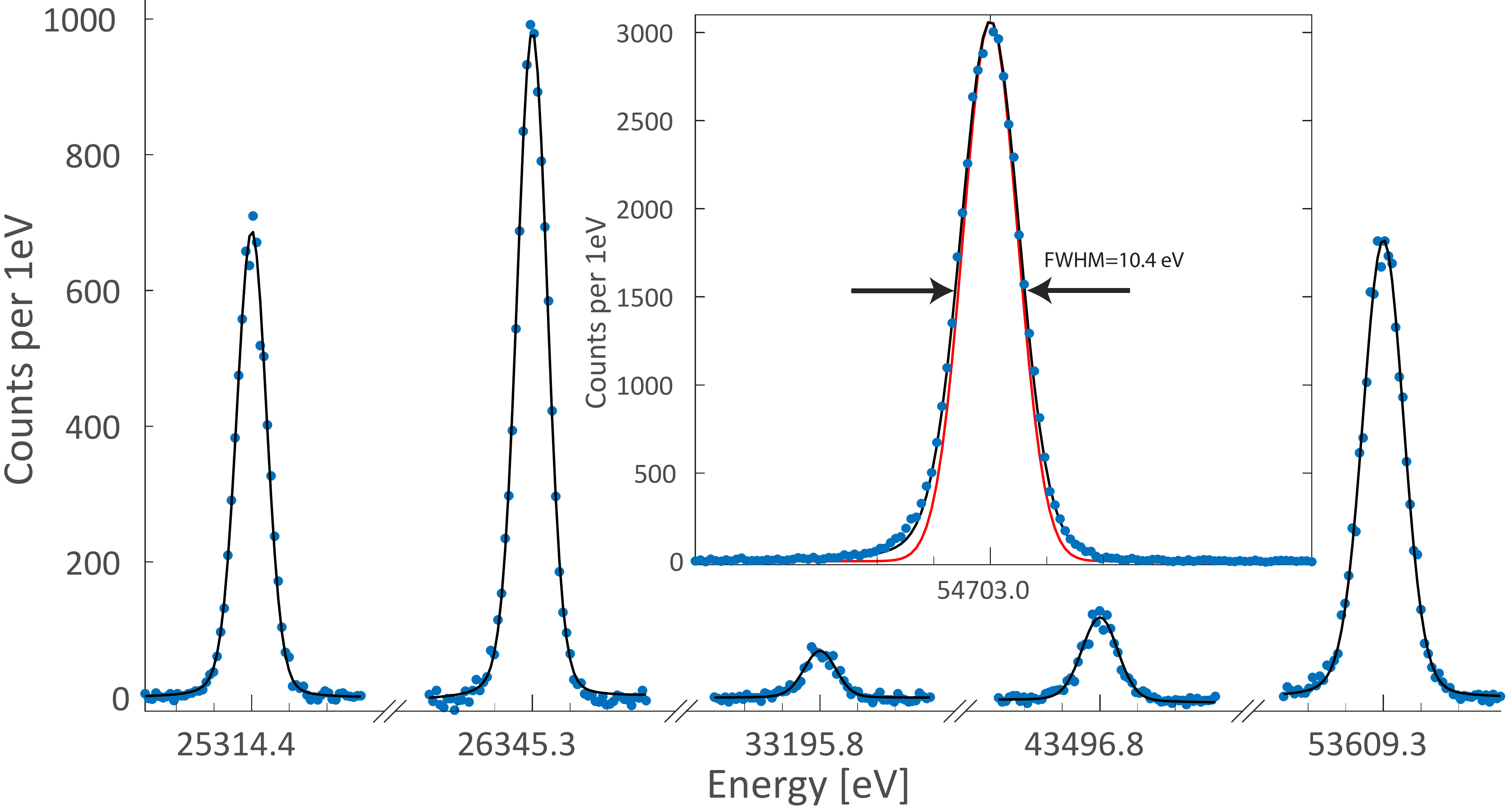}
\caption{Asymmetric Voigt fits of the calibration lines. The inset shows the comparison of the Gaussian (red) and asymmetric Voigt fit (black). The spacing between minor ticks is \SI{5}{\electronvolt}.}
\label{voigt}
\end{figure}

\cleardoublepage
\textbf{Error calculation summary}
\begin{table}[ht]
\centering
\setlength{\tabcolsep}{9pt} 
\begin{tabular}{cccccccc}
\hline
                    & Value & Total error & Statistical & Branching ratio & Np concentration & Fluctuation   &  \\ \hline
4-lines subtraction & 8.10  & 0.17        & 0.12        & 0.03             & \textbf{0.12}    & n/a           &  \\
3-lines subtraction & 8.08  & 1.34        & 0.25        & 0.02             & n/a              & \textbf{1.32} &  \\
Absolute energy     & 7.79  & 0.76        & 0.02        & 0.02             & n/a              & \textbf{0.76} &  \\
Lineshape           & 7.84  & 0.29        & 0.20        & \textbf{0.21}   & n/a              & n/a           &  \\ \hline
\end{tabular}
\caption{Summary of the various contributions to the uncertainty for each scheme to derive the isomer energy. The most dominant contribution for each scheme is marked in bold.}
\label{Table 1}
\end{table}
\end{center}
\end{widetext}
\bibliography{supplement.bib}